\documentclass[preprint,aps,prx,superscriptaddress,nofootinbib,floatfix]{revtex4-2}

\usepackage{adjustbox} % in preamble
\usepackage[english]{babel}
\usepackage{graphicx}
\usepackage{dcolumn}
\usepackage{bm}
\usepackage{amsmath, amssymb, amsfonts,units}
\usepackage{color}
\usepackage{xcolor}
\usepackage{float}
\usepackage{gensymb}
\usepackage{placeins}
\usepackage{multirow}
\usepackage{ulem}
\usepackage{bbm}
\usepackage[hyperfootnotes=false]{hyperref}
\usepackage[left]{lineno} % optional for line numbers
\usepackage{booktabs}    % needed for \toprule etc.
\usepackage{multirow}   
\usepackage{array}

\def\ideal{Co$_{1/3}$TaS$_{2}$}
\def\underdoped{Co$_{0.32}$TaS$_{2}$}
\renewcommand{\arraystretch}{1.2}
\begin{document}
%\linenumbers

\title{Discovery of the (2+1)\textbf{Q} Order in Co$_{0.32}$TaS$_2$}

\title{Anisotropic Multi-\textbf{Q} Order in Co$_{x}$TaS$_2$ }

\author{Jonathon Kruppe}
\affiliation{Department of Physics, University of California, Berkeley, California 94720, USA}
\affiliation{Materials Science Division, Lawrence Berkeley National Laboratory, Berkeley, California 94720, USA}
\author{Josue Rodriguez}
\affiliation{Department of Physics, University of California, Berkeley, California 94720, USA}
\affiliation{Materials Science Division, Lawrence Berkeley National Laboratory, Berkeley, California 94720, USA}
\author{Catherine Xu}
\affiliation{Department of Physics, University of California, Berkeley, California 94720, USA}
\affiliation{Materials Science Division, Lawrence Berkeley National Laboratory, Berkeley, California 94720, USA}
\author{James Analytis}
\affiliation{Department of Physics, University of California, Berkeley, California 94720, USA}
\affiliation{Materials Science Division, Lawrence Berkeley National Laboratory, Berkeley, California 94720, USA}
\affiliation{Kavli Energy NanoScience Institute, Berkeley, CA 94720, USA}
\author{Joseph Orenstein}
\affiliation{Department of Physics, University of California, Berkeley, California 94720, USA}
\affiliation{Materials Science Division, Lawrence Berkeley National Laboratory, Berkeley, California 94720, USA}
\author{Veronika Sunko}
\affiliation{Department of Physics, University of California, Berkeley, California 94720, USA}
\affiliation{Materials Science Division, Lawrence Berkeley National Laboratory, Berkeley, California 94720, USA}
\affiliation {Institute of Science and Technology Austria, Am Campus 1, 3400 Klosterneuburg, Austria }

\begin{abstract}

The cobalt-intercalated transition metal dichalcogenide Co$_x$TaS$_2$ hosts a rich landscape of magnetic phases that depend sensitively on $x$. While the stoichiometric compound with $x=1/3$ exhibits a single magnetic transition, samples with $x\leq 0.325$ display two transitions with an anomalous Hall effect (AHE) emerging in the lower temperature phase. Here, we resolve the spin structure in each phase by employing a suite of magneto-optical probes that include the discovery of anomalous magneto-birefringence -- a spontaneous time-reversal sensitive rotation of the principal optic axes. A symmetry-based analysis identifies the AHE-active phase as an anisotropic (2+1)\textbf{Q} state, in which magnetic modulation at one wavevector (\textbf{Q}) differs in symmetry from that at the remaining two. The (2+1)\textbf{Q} state naturally exhibits scalar spin chirality as a mechanism for the AHE and expands the classification of multi-Q magnetic phases.

\end{abstract}

\maketitle

\section{Introduction}

Realizing macroscopic signatures of time-reversal symmetry (TRS) breaking in antiferromagnets (AFMs) with vanishing net magnetization is an important goal of both applied and fundamental research. Encoding information in time-reversed states of AFMs - in which coupling to stray fields is naturally suppressed - offers a compelling advantage for emerging magnetic computing platforms~\cite{jungwirth_antiferromagnetic_2016,baltz_antiferromagnetic_2018,wadley_electrical_2016,nair_electrical_2020,maniv_exchange_2021,maniv_antiferromagnetic_2021,haley_long-range_2023,qi_antiferromagnetic_2024,cheong_altermagnetism_2024}. Recent theoretical work has made great progress in identifying the classes of AFM order that can manifest TRS-breaking responses such as the anomalous Hall effect (AHE)~\cite{chen_anomalous_2014,Kubler2014,naka_anomalous_2020,liu_intrinsic_2021,naka_anomalous_2022,nguyen_ab_2023,sato_altermagnetic_2024,cheong_altermagnetism_2024,cheong_emergent_2024,cheong_altermagnetism_2025}, prompting searches for materials that exemplify these classes~\cite{surgers_large_2014,nakatsuji_large_2015,suzuki_large_2016,tenasini_giant_2020,nayak_large_2016,smejkal_anomalous_2022,wang_anomalous_2023}.

In transition metal dichalcogenides (TMDs) intercalated with magnetic ions, chirality, geometric frustration, and spin-orbit interaction generate AFM states that exhibit TRS-breaking~\cite{tenasini_giant_2020,park_field-tunable_2022,takagi_spontaneous_2023,park_tetrahedral_2023,ray_zero-field_2025,mayoh_giant_2022,park_contrasting_2025}. However, linking TRS-breaking with specific forms of magnetic order in these materials is challenging, particularly because small changes in the intercalant concentration, $x$, can result in dramatically different properties~\cite{wu_highly_2022,park_composition_2024}. This issue is exemplified by $\text{Co}_{x} \text{TaS}_2$, with $x$ in the vicinity of $1/3$. The stoichiometric compound ($x = 1/3$) exhibits a single magnetic transition, and no AHE~\cite{park_composition_2024,parkin_3_1980}. In contrast, samples with $x < 0.325$ exhibit two magnetic transitions (at $T_{N1}$ and $T_{N2}$), with the AHE appearing in the low-temperature $T_{N2}$ ground state~\cite{park_tetrahedral_2023,park_composition_2024}.

While elastic neutron scattering (NS) is a powerful probe of magnetic states, it often cannot uniquely determine the order. This is particularly true when NS reveals multiple symmetry-related spin-order wavevectors (\textbf{Q}). As NS averages over large sample volumes it is difficult to distinguish a coherent multi-\textbf{Q} structure from an incoherent superposition of single \textbf{Q} domains. To try to resolve this ambiguity, recent studies have combined NS with the symmetry constraints imposed by the AHE. Applied to $\text{Co}_x\text{TaS}_2$, this analysis has led to the proposal that the phase that exhibits AHE is a coherent superposition of spin order with modulation at three symmetry-related wave vectors ($3\textbf{Q}$ order)~\cite{takagi_spontaneous_2023,park_tetrahedral_2023}. The proposed non-coplanar spin structure exhibits scalar spin chiral order, which is regarded as a source of the AHE~\cite{taguchi2001spin,verma_unified_2022}, and possesses 3-fold rotational symmetry.

Here we show that the proposed 3\textbf{Q} order is incompatible with optical measurements that show that 3-fold symmetry is broken for all $T<T_{N1}$. Birefringence, which is the optical signature of rotational symmetry breaking, onsets at $T_{N1}$, and increases in magnitude rapidly below $T_{N2}$. Furthermore, we show that for $T<T_{N2}$, the spin order is uniquely specified by the observation of a rotation of the birefringent axes whose direction follows the sign of the magneto-optic Kerr effect. Our symmetry analysis points to a form of anisotropic non-coplanar order, which we term (2+1)\textbf{Q} order, as the only state consistent with all experimental observations. In the (2+1)\textbf{Q} state the magnetic modulation at one of the three wavevectors (1\textbf{Q}) has different symmetry from that at the remaining two wavevectors (2\textbf{Q}), making this state fundamentally distinct from the previously proposed 3\textbf{Q} structure. The (2+1)\textbf{Q} structure exhibits scalar spin-chirality (SSC), extending the class of magnetic textures known to support this mechanism for AHE.

\begin{figure*}[t!]
\includegraphics[width=\textwidth]{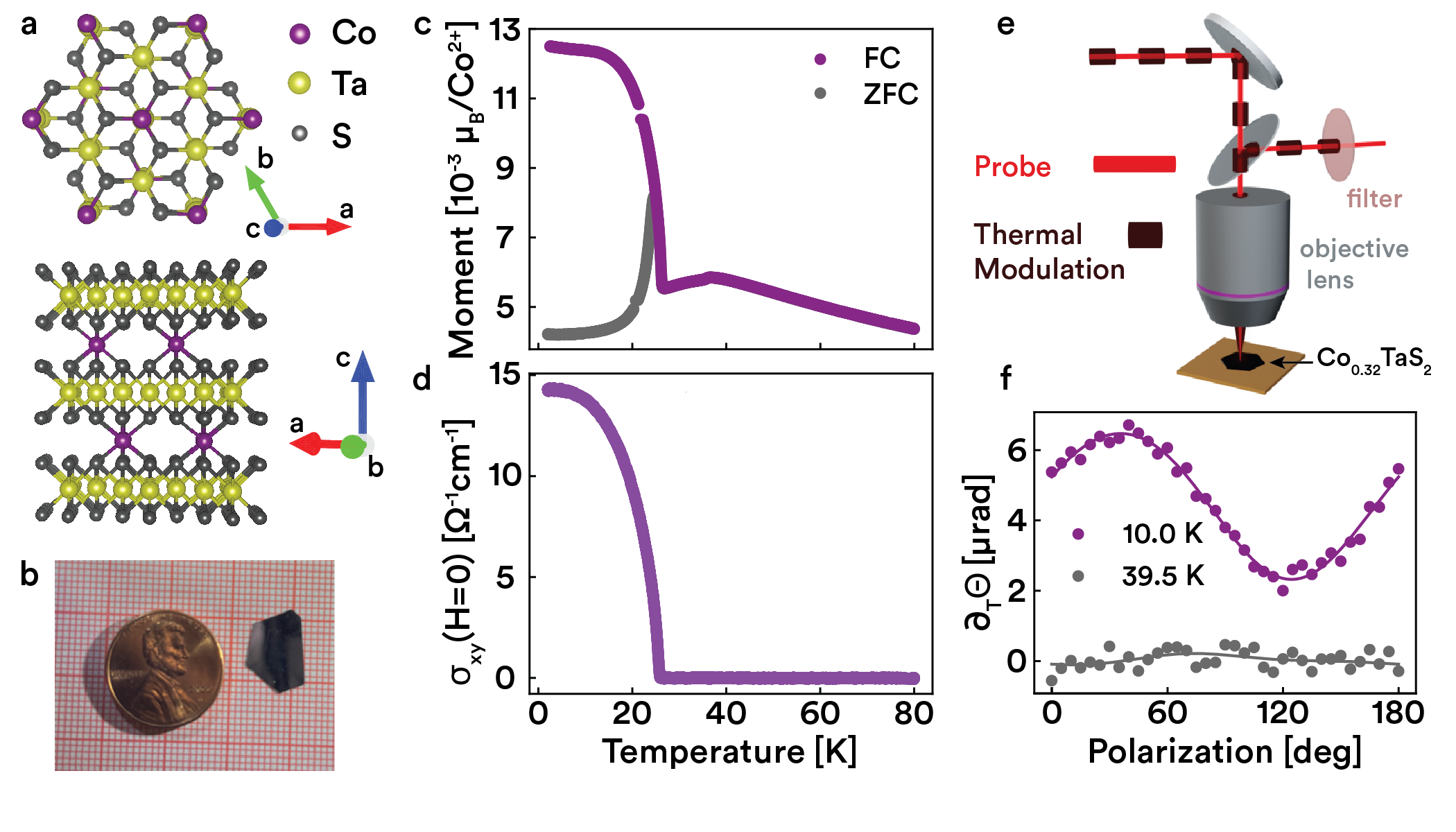}
\caption{(a) Crystal structure of $\text{Co}_{0.32} \text{TaS}_2 $ top and side view. (b) Image of large single crystals. (c) Zero-field and field-cooled magnetization as a function of temperature. The temperature dependence of the moment reveals two transitions at 36 K (HT phase) and 26 K (LT phase). (d) Anomalous Hall conductivity turns on in the LT phase. (e) Schematic of optical setup for measuring polarization rotation ($\Theta$) as a function of incident polarization ($\varphi$): a probe laser (HeNe, 633 nm, 150 $ \mu W $) is focused onto the sample via an objective lens. A second optically chopped laser (780 nm, 20 kHz, 50 $ \mu W$) is spatially overlapped with the probe to modulate the sample temperature at the chopping frequency. The thermal-modulation beam is filtered out before detection. (f) Thermally-modulated polarization rotation vs polarization in the LT phase (10 K) and in the paramagnetic phase (39 K), fitted to S10. The 10 K data reveal that LT phase not only breaks time-reversal symmetry (offset) but also $ C_3 $ symmetry (sinusoidal variation), which was not previously reported.
\label{fig:Fig1}
}
\end{figure*}

\section{Sample growth and characterization}

Large single crystals of $\text{Co}_{0.32} \text{TaS}_2 $ (see Fig.~\ref{fig:Fig1}a,b) were grown by chemical vapor transport and the intercalation was precisely determined by energy-dispersive x-ray spectroscopy (SI~\cite{kruppeSI2025}). The crystal structure, in which the intercalated Co ions form a triangular lattice, is shown in Fig.~\ref{fig:Fig1}a. The temperature dependence of the magnetization (Fig.~\ref{fig:Fig1}c) and the AHE (Fig.~\ref{fig:Fig1}d) agree with results previously reported on samples in this intercalation range~\cite{park_tetrahedral_2023,park_composition_2024}. The magnetization reveals two magnetic transitions as a function of temperature, at $T_{N1}=36 \text{K}$ and $T_{N2}=26 \text{K}$. Both high temperature (HT, $T_{N1}>T>T_{N2}$) and low temperature (LT, $T<T_{N2}$) magnetic phases are antiferromagnetic. The onset of AHE coincides with the appearance of the LT phase, which also exhibits a weak ferromagnetic moment revealed through the difference in magnetization measured upon warming after cooling in zero field and cooling in a field of $H=0.3 $ T.

\section{Optical measurements}

We now turn to optical experiments which probe the broken symmetries in each of the two magnetic phases. We study a sample from the same batch as those used to establish the bulk properties. Using the setup shown in Fig.~\ref{fig:Fig1}e, we illuminate the sample with linearly polarized laser light, and measure the change of polarization angle upon reflection ($\Theta$) as a function of incident light polarization ($\varphi$). This experiment is sensitive to breaking of time-reversal symmetry (TRS) and the three-fold rotational symmetry around the $z$ axis ($C_{3z}$). Broken TRS yields the magneto-optical Kerr effect (MOKE), i.e. a rotation of polarization angle upon reflection that is independent of incident polarization angle ($\Theta(\varphi)=\theta_K$). Broken $C_{3z}$ - symmetry allows for optical anisotropy, so-called birefringence, which results in a $\Theta$ that depends sinusoidally on $\varphi$, and vanishes when $\varphi$ coincides with principal optical axes (SI~\cite{kruppeSI2025}). The measured polarization rotation is given by: 

\begin{equation}
\Theta (\varphi,T) = \theta_B(T)\sin{\left[2(\varphi +\varphi_0(T))\right]} +\theta_K(T)
\label{eq:corotation}
\end{equation}

\noindent with $ \varphi $ being the incident polarization, $ \varphi_0 $ the orientation of the optical principal axes, $ \Theta $ the polarization rotation upon reflection, $ \theta_B $ the amplitude of birefringence, $ \theta_K $ the amplitude of the Kerr effect, and $ T $ the temperature. We enhance the measurement sensitivity by thermal modulation provided by a second overlapping laser beam. Synchronous detection of $\Theta$ at the modulation frequency yields a signal proportional to its temperature derivative, $\partial_T \Theta$~\cite{little_three-state_2020,sun_mapping_2021,sunko_spin-carrier_2023, donoway_multimodal_2024,kruppeSI2025}.

In Fig.~\ref{fig:Fig1}e we compare $\partial_T\Theta(\varphi)$ vs. $\varphi$ measured at a temperature above both transitions (39.5 K) and in the LT phase (10 K). The higher temperature measurement shows neither an offset nor a sinusoidal variation, whereas in the LT phase both appear. MOKE is expected, since it has the same symmetry as AHE, which is observed in the LT phase (Fig.~\ref{fig:Fig1}d). However, birefringence is incompatible with the proposed $C_{3z}$-symmetric 3\textbf{Q} structure~\cite{takagi_spontaneous_2023,park_tetrahedral_2023}. This measurement therefore shows that the current understanding of the LT magnetic structure is incomplete.

%%%%%%%%%%%%%%%%%%%%%%%%%%% figure 2 %%%%%%%%%%%%%%%%%%%%%%%%%%%%%%%%
\begin{figure}[t!]
\includegraphics[width=0.5\columnwidth]{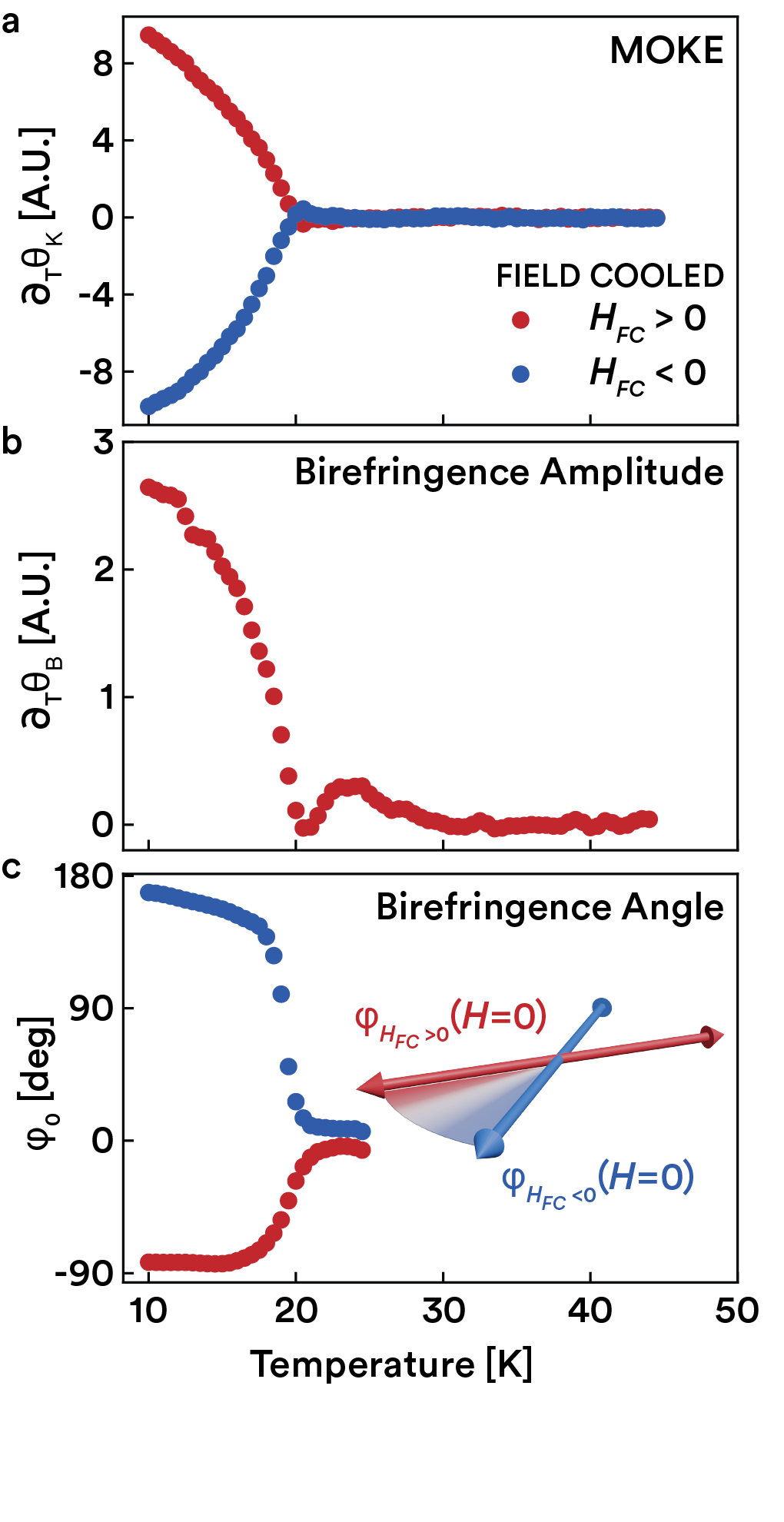}
\caption{Temperature dependence of thermally-modulated (a) MOKE, (b) birefringence and (c) principal axis orientation ($ \varphi_0 $). Samples field cooled in $ H_{FC} \pm  500 \text{mT}$ field, measurements are in zero-field. Values are extracted from rotation measurements according to the procedure described in~\cite{kruppeSI2025}. (a) MOKE onsets in LT phase, with sign set by the sign of $H_{FC}$. (b) Birefringence is non-zero in both phases. (c) $ \varphi_0 $ rotates in the LT phase depending on the sign $ H_{FC}$; (inset) AMB: principal optic axes rotate in response to intrinsic time-reversal breaking order parameter.
\label{fig:Fig2}
}
\end{figure}
%%%%%%%%%%%%%%%%%%%%%%%%%%% figure 2 %%%%%%%%%%%%%%%%%%%%%%%%%%%%%%%%

In order to study the evolution of the symmetry breaking, we measure $\partial_T\Theta $ as a function of temperature. The sample is field-cooled (FC), and measured upon warming in zero field. We refer to this protocol as $\pm$FC, for positive and negative cooling fields, respectively. The temperature derivative of the Kerr rotation, $\partial_T\theta_K$, can be extracted directly from the polarization-independent component of $\partial_T\Theta$. Extracting the temperature dependence of the birefringence requires additional analysis, because the sinusoidal component of $\partial_T\Theta $ depends on $\theta_B(T)$, $\varphi_0(T)$, and their derivatives. Although straightforward, this analysis requires a detailed description, which we provide in the SI~\cite{kruppeSI2025}.

In Fig.~\ref{fig:Fig2}(a, b) we plot $\partial_T\theta_K$ and $\partial_T\theta_B$, respectively, as a function of temperature, showing that MOKE and birefringence onset at different temperatures: $T'_{N1}= 30 \text{K}$ for birefringence and $T'_{N2}= 20 \text{K}$ for MOKE. While $T'_{N1}$ and $T'_{N2}$ are not identical to $T_{N1}$ and $T_{N2}$ deduced from magnetization and AHE measurements (likely the result of imperfect thermalization in the optical cryostat) their separation in temperature is ($T'_{N1}-T'_{N2}\approx T_{N1}-T_{N2}$). We therefore assign $T'_{N1}$ and $T'_{N2}$ as the transition temperatures of the HT and LT phases, respectively. MOKE (Fig.~\ref{fig:Fig2}a) behaves as expected based on previous findings~\cite{park_field-tunable_2022,takagi_spontaneous_2023,park_tetrahedral_2023} and general symmetry considerations: it is non-zero only in the LT phase, and its sign is selected by the sign of the field in which the sample is cooled (Fig.~\ref{fig:Fig2}a). In contrast, the non-zero birefringence provides fundamentally new information about the magnetic structures: both the LT and the HT phase break rotational symmetry. We note that $\partial_T\theta_B$ is qualitatively the same for $\pm$FC, so for clarity we show only the $+$FC measurement.

Although birefringence is non-zero in both HT and LT phase, its properties in the two phases are quite different. In the HT phase the birefringence angle is approximately independent of both the temperature and the cooling protocol. At the transition to the LT phase, the angle becomes abruptly temperature-dependent. $\partial_T\theta_B$ exhibits a minimum at $T^\prime_{N2}$, followed by a rapid rotation of the birefringence angle $\varphi_0$ (Fig.~\ref{fig:Fig2}c). Moreover, the sign of the change of $\varphi_0$ depends on the sign of the cooling field, with the difference between the $\pm$ FC angles increasing with decreasing temperature.

The rotation of the principal axes in the LT phase is reminiscent of the off-diagonal linear magneto-birefringence (LMB), in which rotation is induced by an applied field~\cite{eremenko_birefringence_1980,eremenko_magneto-optics_1987,kharchenko_linear_1994,sunko_spin-carrier_2023,donoway_multimodal_2024}. However, here the rotation appears in zero-field as a consequence of spontaneous time-reversal breaking, whose sign is trained by the field applied during cooling. We therefore refer to the spontaneous rotation as ``anomalous" by analogy with the relationship between the conventional Hall effect and the AHE, and denote the effect as the anomalous magneto-birefringence (AMB). AMB shares the same symmetry as LMB, and its observation introduces constraints on the candidate magnetic structures beyond those posed by birefringence and MOKE.

\section{Symmetry Analysis Reveals the (2+1)\textbf{Q} Ground State}

%%%%%%%%%%%%%%%%%%%%%%%%%%% figure 3 %%%%%%%%%%%%%%%%%%%%%%%%%%%%%%%%
\begin{figure*}[t!]
\includegraphics*[width=\textwidth]{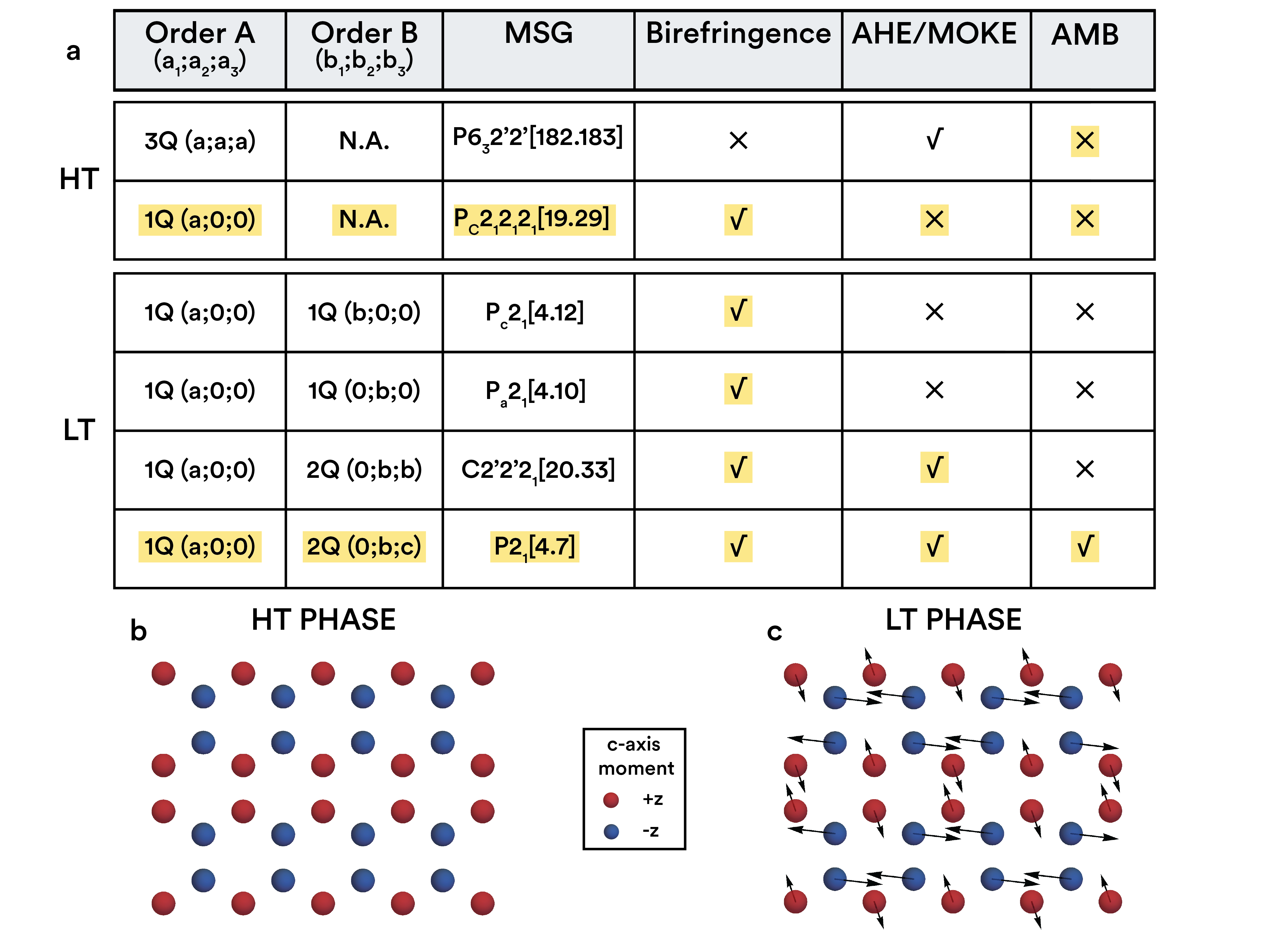}
\caption{(a) Results of symmetry analysis combining neutron scattering and optics for both magnetic phases. HT phase matches a 1\textbf{Q} structure [19.29]. In the LT phase, an additional 2\textbf{Q} structure allows for all the observed effects yielding MSG 4.7. Visualizations of the structures in the (b) HT phase with out-of-plane order and (c) LT phase with (2+1)Q order.
\label{fig:Fig3}
}
\end{figure*}
%%%%%%%%%%%%%%%%%%%%%%%%%%% figure 3 %%%%%%%%%%%%%%%%%%%%%%%%%%%%%%%%
To determine the magnetic structures consistent with all experimental observations, we follow a multimodal approach~\cite{donoway_multimodal_2024}: first identifying which magnetic space groups (MSGs) are compatible with neutron scattering~\cite{takagi_spontaneous_2023,park_tetrahedral_2023}, and then determining which of these allow the observed optical effects. A summary of the possible magnetic order parameters, their corresponding MSGs and optical properties is provided in Fig.~\ref{fig:Fig3}a. This analysis reveals an anisotropic non-coplanar multi-\textbf{Q} order as the ground state.

The starting point of our analysis is the usual assumption that $\text{Co}_{0.32}\text{TaS}_2$ belongs to the same space group as stoichiometric $\text{Co}_{1/3}\text{TaS}_2$. Neutron scattering revealed that spins in the HT phase are oriented perpendicular to the Co planes, and that the only irreducible representation (irrep) consistent with the data is $\mathrm{mM}_2$~\cite{takagi_spontaneous_2023,kruppeSI2025}. At the transition into the LT phase the scattering intensity associated with $\mathrm{mM}_2$ changes smoothly. However, a new set of scattering peaks appears, corresponding to in-plane spin order and the irrep $\mathrm{mM}_4$~\cite{takagi_spontaneous_2023,kruppeSI2025}. 

Both $\mathrm{mM}_2$ and $\mathrm{mM}_4$ are three-dimensional irreducible representations, so the corresponding magnetic structures are linear combinations of three basis functions, with coefficients $(a_1; a_2; a_3)$ for $\mathrm{mM}_2$ and $(b_1; b_2; b_3)$ for $\mathrm{mM}_4$. These basis functions correspond to spin order at the three symmetry-equivalent \textbf{Q} vectors. For example, 1\textbf{Q} and 3\textbf{Q} structures from $\mathrm{mM}_2$ correspond to $(a;0;0)$ and $(a;a;a)$, respectively. In the LT phase, the combined order parameter takes the form $(a_1; a_2; a_3 \mid b_1; b_2; b_3)$, where the first three coefficients describe the $\mathrm{mM}_2$ order parameter and the latter three the $\mathrm{mM}_4$ one. The MSG is determined by which $a_i$ and $b_i$ are non-zero.

To establish the magnetic structure in the two phases, we examine which of the MSGs are compatible with the observed birefringence, MOKE (which shares the symmetry of the AHE), and AMB. We perform this analysis using the MAGNDATA tool of the Bilbao Crystallographic Server~\cite{Gallego:lk5043}, as detailed in the SI~\cite{kruppeSI2025}. For the HT phase we analyze structures associated with $\mathrm{mM}_2$, and find that the 1\textbf{Q} structure described by $(a;0;0)$ is consistent with all the observations (MSG 19.29, illustrated in Fig.~\ref{fig:Fig3}b), while the 3\textbf{Q} structure is not. The conclusion that the HT phase has 1\textbf{Q} order was reached previously in Ref.~\cite{park_tetrahedral_2023}.

The $(a;0;0)$ order of the $\mathrm{mM}2$ irrep persists below $T_{N2}$, as the associated scattering remains unchanged through the transition~\cite{takagi_spontaneous_2023}. We therefore search for an $\mathrm{mM}_4$ order compatible with the existing $\mathrm{mM}_2$ structure, and identify four such $\mathrm{mM}_4$ order parameters, listed  in Fig.~\ref{fig:Fig3}a, along with the corresponding MSGs and allowed optical effects. Since rotational symmetry is already broken by the $(a;0;0)$ order, all four candidate structures allow birefringence. However, the analysis of corresponding MSGs reveals that only those in which $\mathrm{mM}_4$ order has two components allow MOKE.  The combined order parameter therefore must take the form $(a; 0; 0 \mid 0; b; c)$.

If only birefringence and MOKE data were available, structures with $b = c$ could not be distinguished from those with $b \ne c$. However, the $b = c$ structure is symmetric under a combination of time-reversal and a two-fold rotation about an in-plane axis. This symmetry prohibits the off-diagonal LMB~\cite{sunko_spin-carrier_2023}, and therefore the observed AMB as well. The combination of neutron scattering and the three optical techniques thus identifies $(a; 0; 0 \mid 0; b; c)$ (MSG 4.7), illustrated in Fig.~\ref{fig:Fig3}c, as the only structure consistent with all available data.

The above analysis reveals a magnetic structure for the LT phase that differs from previous reports~\cite{takagi_spontaneous_2023,park_tetrahedral_2023,kim_electrical_2024,park_composition_2024}. Although three \textbf{Q} values are active, the structure is not rotationally symmetric: the order at one of the \textbf{Q} vectors belongs to the $\mathrm{mM}_2$ irrep, while the order at the other two belongs to $\mathrm{mM}_4$. We refer to this form of anisotropic multi-\textbf{Q} order as a (2+1)\textbf{Q} structure. 

The (2+1)\textbf{Q} structure we identify is non-coplanar, and therefore could host SSC. To explore this possibility, we analytically evaluate the SSC-induced fictitious magnetic flux $\bm{b}$ for the combined magnetic order parameter, and find~\cite{kruppeSI2025}:
\begin{equation}
\bm{b} \propto \left( a_2 b_1 b_3 - a_3 b_1 b_2 - a_1 b_2 b_3 \right) \bm{\hat{z}}.
\end{equation}\label{eq:SSC}

\noindent The (2+1)\textbf{Q} structure, which exhibits non-zero $a_1$, $b_2$ and $b_3$ and vanishing other coefficients, therefore supports SSC for all parameter values. Unlike the rotationally symmetric 3\textbf{Q} structure, it does not require condensing all six components of the order parameter. Further, the anisotropy of the (2+1)\textbf{Q} structure means that uniaxial strain can couple to it, creating the possibility to tune SSC via strain.

\section{Conclusions and Outlook}

In this work, we combined a trio of symmetry-sensitive magneto-optical probes with existing neutron scattering measurements to identify the magnetic structure of $\text{Co}_{0.32}\text{TaS}_2$, which we refer to as (2+1)\textbf{Q} order. In this anisotropic multi-\textbf{Q} phase, the spin modulation at one of the three wavevectors has different symmetry from that at the remaining two, making this state fundamentally distinct from the previously proposed 3\textbf{Q} structure. Like the 3\textbf{Q} phase, (2+1)\textbf{Q} structure manifests scalar spin chiral order, consistent with the observation of the AHE.

The key to our identification of the (2+1)\textbf{Q} phase was the discovery that the principal optic axes that appear at $T_{N1}$ begin to rotate as the temperature is lowered through $T_{N2}$. Just as the Kerr effect at optical frequency is linked to the AHE at $\omega=0$, we expect that the anomalous magneto-birefringence that we have discovered should have an associated $dc$-transport effect: the appearance of symmetric off-diagonal components in the resistivity tensor that are linearly related to the AHE~\cite{sunko_linear_2023}. Future investigations are needed to confirm this transport  effect and explore its possible link to scalar spin chirality.

Finally, our work brings into focus the extreme sensitivity of the magnetic structure in $\text{Co}_{x}\text{TaS}_2$ to small departures, $\delta x$, from $x=1/3$. Notably, the (2+1)\textbf{Q} state does not appear in stoichiometric $\text{Co}_{1/3}\text{TaS}_2$, which instead exhibits a single transition into a phase characterized by $\bm{Q}=(1/3,0,0)$~\cite{park_composition_2024,parkin_magnetic_1983}. Controlling the properties of $\text{Co}_{x}\text{TaS}_2$, and indeed the whole family of intercalated TMDs, requires understanding the role of small departures  from stoichiometry. While spatially averaged properties such as the c-axis lattice constant are found to vary smoothly with $x$, it difficult to see how this can account for the singular dependence on $\delta x$.  Instead we suggest that the transition to states with complex multi-\textbf{Q} order as observed here are stabilized by the presence of vacancies in the Co superlattice, which can promote higher-order and multi-spin coupling \cite{maryasin_triangular_2013,maryasin_collective_2015} and long-range interactions ~\cite{zhitomirsky2025defect}.

\begin{acknowledgments}
We thank Linda Ye and Yue Sun for helpful discussion. Experimental and theoretical work at LBNL and UC Berkeley was funded by the Quantum Materials (KC2202) program under the U.S. Department of Energy, Office of Science, Office of Basic Energy Sciences, Materials Sciences and Engineering Division under Contract No. DE-AC02-05CH11231. JK received support from the National Science Foundation Graduate Research Fellowship Program under Grant No. 2146752. Any opinions, findings, and conclusions or recommendations expressed in this material are those of the author(s) and do not necessarily reflect the views of the National Science Foundation. V.S. and J.O. received support from the Gordon and Betty Moore Foundation’s EPiQS Initiative through Grant GBMF4537 to J.O. at UC Berkeley.
\end{acknowledgments}

%************************************************************************
\bibliographystyle{unsrt}
\bibliography{maintextreferences}
%************************************************************************

\newpage

% reset things for Methods$
\counterwithout{equation}{section}
\renewcommand\theequation{M\arabic{equation}}
\renewcommand\thefigure{M\arabic{figure}}
\renewcommand\thetable{M\arabic{table}}
\renewcommand\thesection{M\arabic{section}}
\renewcommand\bibnumfmt[1]{[M#1]}
\setcounter{equation}{0}
\setcounter{figure}{0}
\setcounter{enumiv}{0}

\section*{Methods}

\subsection{Sample Growth}

High quality single crystals of Co$_x$TaS$_2$ were grown by a two-step procedure. First, a precursor was prepared. The elements were combined in a ratio Co:Ta:S  ($x$:1.0:2.0). The powders were loaded in alumina crucibles and sealed in quartz tubes under a partial pressure (200 torr) of Argon gas. The tube was heated to 900\degree C and kept there for 10 days. The furnace was then shut off and allowed to cool naturally. This reaction yields a free-flowing shiny powder of polycrystals that was ground with a mortar and pestle separately for each batch.
Second, 1g of precursor was loaded with 10mg/cm$^3$ of iodine in a 10”  long quartz tube, evacuated, and placed in a horizontal two-zone furnace. The precursor and iodine were placed in zone 1 and the other end of the tube (the growth zone) were in zone 2. Both zones were heated to 850\degree C for 6 hours to encourage nucleation. Then, zone 1 was raised at 950\degree C while zone 2 was reduced to 850\degree C. This condition was maintained for 10 days. The furnace was then shut off and allowed to cool naturally. Shiny hexagonal crystals up to 1cm in lateral length were collected from the cold zone. The crystals are easily exfoliated with a scalpel or scotch tape.

\subsection{Bulk Characterization}

The Hall effect and magnetization measurements were performed in a QuantumDesign Physical Property Measurement System (PPMS). For the Hall effect data, a current of 5 milliamps was sent through the sample and modulated at a frequency of 277 Hz. The resistivity was measured via a lock-in detection scheme by demodulating the response at the modulation frequency. The sample was field-cooled in ($+$/$-$) 9T. At 2K the field was set to zero and Hall voltage was measured from 2K to 80K. The Hall signal was anti-symmetrized to remove any longitudinal component and the longitudinal component was symmetrized to remove any transverse component. The zero field Hall conductivity was calculated by

\begin{equation}
\sigma_{xy} = -\frac{\rho_{xy}}{\rho^2_{xx} + \rho^2_{yy}}
\end{equation}

The magnetization measurements were done with the Vibrating Sample Magnetometer (VSM) module of the PPMS. A 12.3 mg sample was glued to a quartz rod and mechanically oscillated through a pickup coil. A 300 mT field was applied to enhance the response of the crystal while measuring. This field does not alter the magnetic state of the sample.

\subsection{Optical setup}
A schematic showing the optical setup used in these experiments is shown in \ref{fig:setup_supp}. A probe beam (633 nm, 150 $ \mu W$) is sent through an initial half-wave-plate (HWP 1) which fixes the incident polarization ($ \varphi $) of the probe. The probe is focused onto the sample by an objective lens. After reflecting off the sample, the probe is sent through a second half-wave plate (HWP 2) which rotates the probe polarization by $45 ^{\circ}$. After HWP 2, the probe is sent through a Wollaston prism which spatially separates horizontally ($E_x $) and vertically ($E_y $) polarized light into two beams. Each of these beams is focused onto a separate photodiode of a balanced photodetector (Thorlabs {\normalfont\#}PDB210A). The output of the balanced detector is proportional to the difference between the intensity incident on each photodiode, $I \sim E_x^2 - E_y^2 $.

By using two waveplates in this fashion, we are measuring the rotation of polarization ($ \Theta $) by the sample at each incident polarization ($ \varphi $); if the sample did not rotate the polarization, then the probe would contain an equal amount of horizontal and vertically polarized light, leading to no signal.

To enhance sensitivity and mitigate spurious contributions arising from the optical anisotropy of the setup itself, we overlap the probe beam with an optically chopped pump beam (780 nm, 50 $ \mu W$, 20 kHZ). This pump beam modulates the temperature of sample at the chopping frequency, and the measured signal is demodulated using a lock-in detection scheme synchronized to the chopper. As shown in Fig.~\ref{fig:setup_supp}, the pump is filtered out prior to detection. Since setup artifacts are insensitive to temperature, their contribution to the measured signal is diminished by this scheme.

% reset things for ED$

\onecolumngrid
\newpage
\newpage
\section*{Extended data}

\counterwithout{equation}{section}
\renewcommand\theequation{ED\arabic{equation}}
\renewcommand\thefigure{ED\arabic{figure}}
\renewcommand\thetable{ED\arabic{table}}
\renewcommand\thesection{ED\arabic{section}}
\renewcommand\bibnumfmt[1]{[ED#1]}
\setcounter{equation}{0}
\setcounter{figure}{0}
\setcounter{enumiv}{0}

\begin{figure*}[h]
\includegraphics*[width=\textwidth]{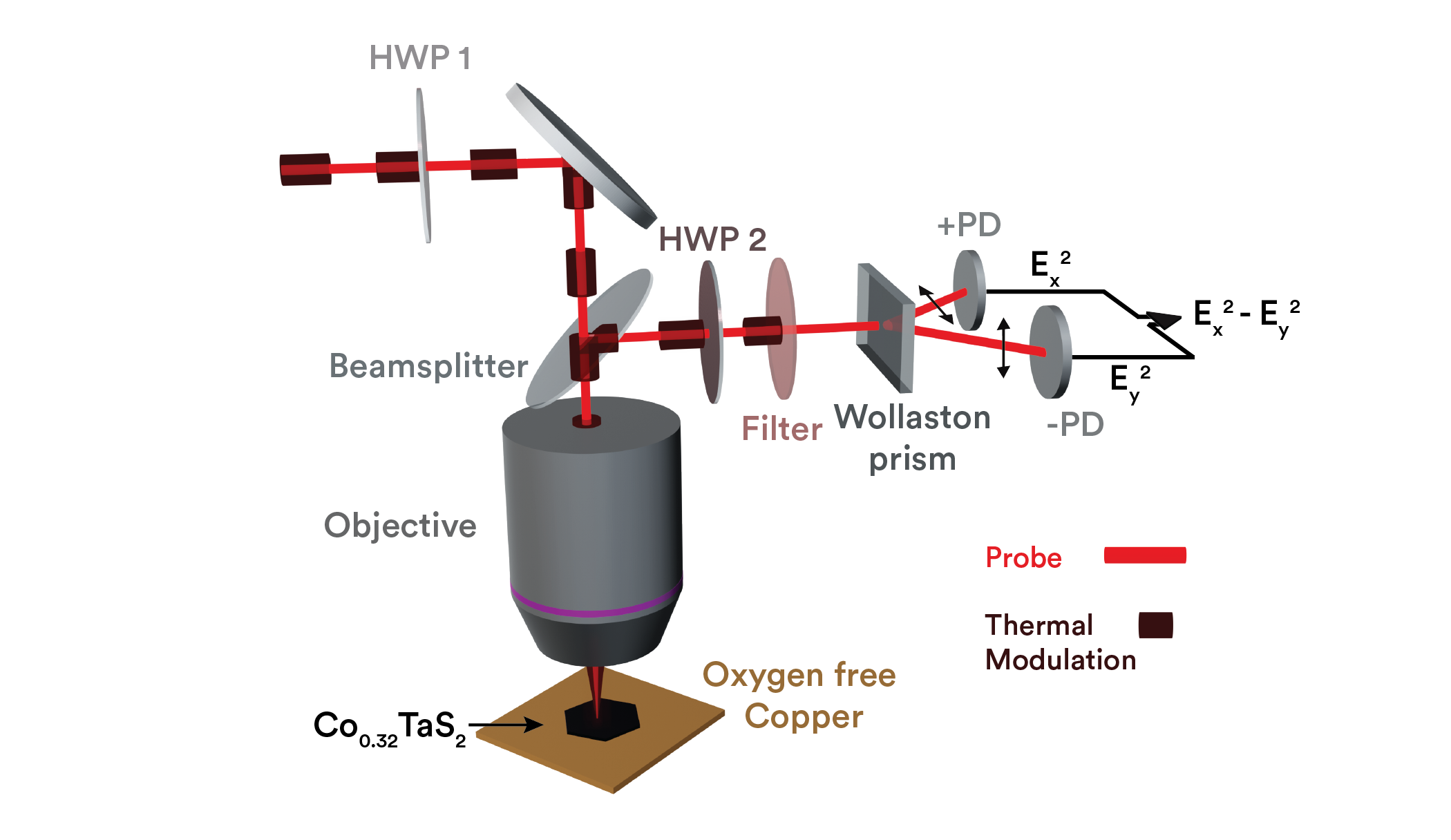}
\caption{Schematic of the setup used for thermally-modulated polarization rotation experiments.
\label{fig:setup_supp}
}
\end{figure*}

% reset things for SI$
\counterwithout{equation}{section}
\renewcommand\theequation{S\arabic{equation}}
\renewcommand\thefigure{S\arabic{figure}}
\renewcommand\thetable{S\arabic{table}}
\renewcommand\thesection{S\arabic{section}}
\renewcommand\bibnumfmt[1]{[S#1]}
\setcounter{equation}{0}
\setcounter{figure}{0}
\setcounter{enumiv}{0}
\setcounter{section}{0}

\newpage
\setcounter{linenumber}{1}   
\section*{Supplementary Information}

\section{Temperature-modulated polarization rotation}

In the supplemental information of Ref.~\cite{sunko_spin-carrier_2023}, the results of such a temperature modulation experiment were modeled using a Jones calculation. We refer to that for further details on the calculation and reproduce here the results relevant to our experiment. The Jones matrix describing the sample is 
\begin{equation}
J_s=\begin{pmatrix}
r + b & k \\
-k & r - b
\end{pmatrix},
\end{equation}

\noindent where $r$ is the reflectivity of the sample, $b$ generates the birefringence and $k$ generates the Magneto-Optic Kerr Effect (MOKE). With no temperature modulation, the expected polarization rotation ($ \Theta $) from the sample is given by~\cite{sunko_spin-carrier_2023}:

\begin{equation}
\Theta(\varphi,T) \equiv \frac{I(\varphi)}{r^2} = -\frac{2b}{r} \text{sin}[2(\varphi + \varphi_0)] - \frac{2k}{r} 
\label{eq:idealsample}
\end{equation}

\noindent where $ I = E_x^2 - E_y^2 $ is the output of the balanced detector and $ \varphi $ the incident polarization. By identifying $ \theta_k = -\frac{2k}{r} $ and $ \theta_B = -\frac{2b}{r} $, we arrive at a description of the polarization rotation induced by the sample:

\begin{equation}
\Theta(\varphi,T) = \theta_B(T) \text{sin}[2(\varphi + \varphi_0(T))] + \theta_K(T) .
\label{eq:idealsample}
\end{equation}

The objective of the following analysis is to describe how we obtain these sample-related parameters $ \theta_k $, $ \theta_B $ and $ \phi_0 $ from our temperature-modulation measurements. 

\subsection{Setup induced artifacts}
A principal goal of thermal-modulation is to suppress setup-induced contributions to polarization rotation. While the optical components can produce temperature-independent polarization rotation on  the order of $10\,\mathrm{mRad}$, their effects are largely eliminated through thermal modulation. However, a setup-related contribution does appear in a thermal modulation experiment if the reflectivity of the sample is temperature dependent. We describe below how it is easily subtracted off, starting with the expression for setup-induced polarization rotation~\cite{sunko_spin-carrier_2023}:

\begin{equation}
\frac{I_{\text{S}}(\varphi)}{r^2} = \epsilon \sin(2\varphi) - \sin^2\left(\frac{\delta}{2}\right) \sin(4\varphi)\label{eq:setup}
\end{equation}

\noindent where $ \epsilon $ is the setup birefringence and $ \delta $ is the setup retardance. In the thermally-modulated experiment, we measure the temperature derivative of the above. The thermally-modulated setup contribution is:

\begin{equation}
\frac{I_{\text{ST}}(\phi)}{r^2} = 
2 \frac{\partial_t r}{r} \left( 
\epsilon \sin(2\varphi) 
- \sin^2\left( \frac{\delta}{2} \right) \sin(4\varphi)
\right).
\label{eq:dsetup}
\end{equation}

\noindent where $ \partial_t r$ is the temperature-modulated reflectivity, and the quantity in parentheses is the unmodulated setup birefringence.

\begin{figure*}[t!]
\includegraphics*[width=\textwidth]{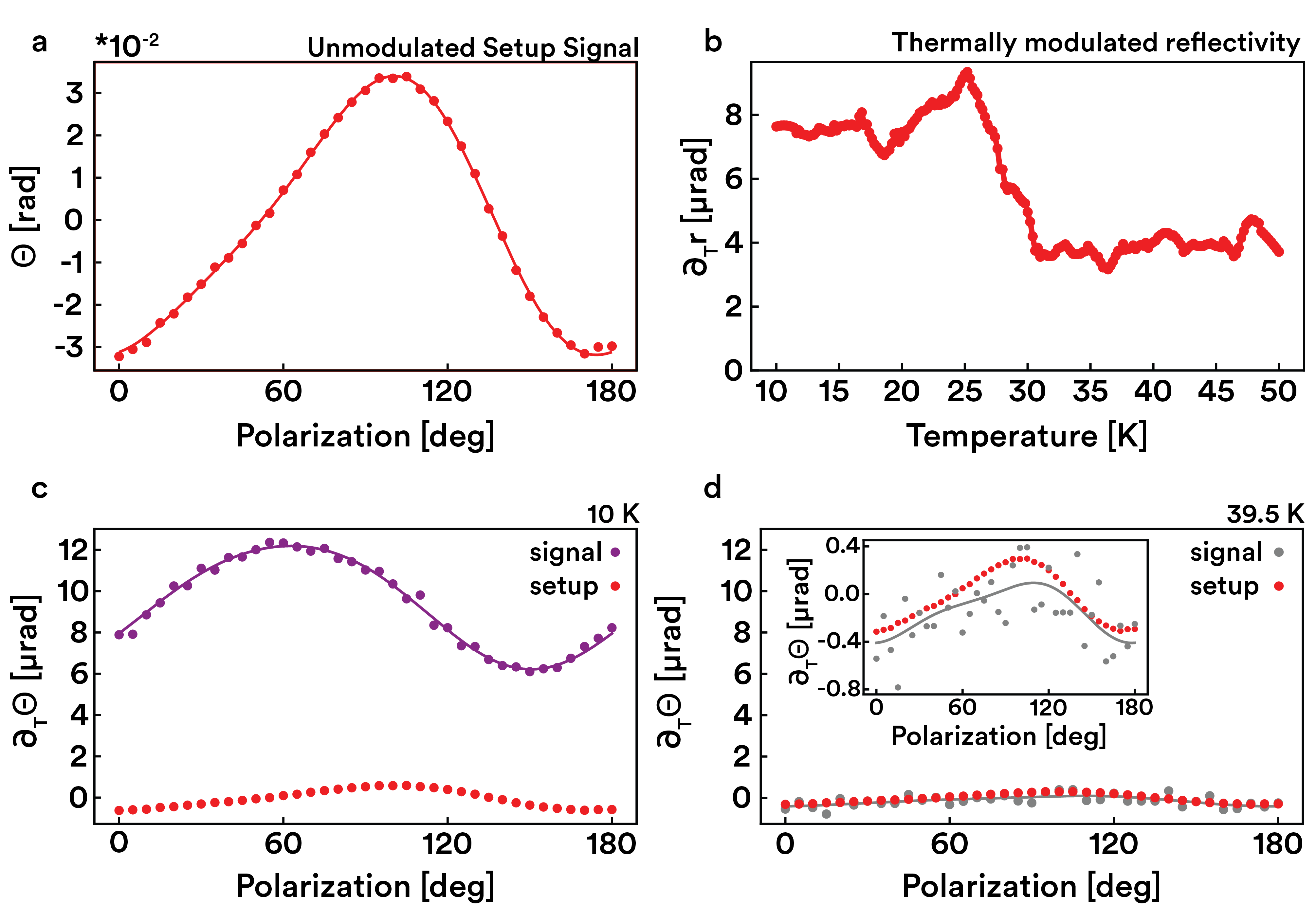}
\caption{Demonstration of setup artifact from thermally-modulated polarization rotation measurements (a) polarization rotation experiment without any modulation. The measured signal purely reflects the setup contribution as any rotation from the sample is negligible in the absence of modulation (b) temperature dependence of the total thermally-modulated reflectance, which can be combined with the unmodulated setup signal in (a) to obtain \ref{eq:dsetup} the setup artifact. Comparison of setup artifact and total thermally-modulated signal at both (c) low temperature and (d) high temperature. The data in (d) are taken well above both magnetic transitions, where the thermally-modulated signal is entirely due to the setup artifact \ref{eq:dsetup}. This is evident from the overlap between the setup and signal curves.
\label{fig:background_sub}
}
\end{figure*}

We can subtract the setup contribution described in \ref{eq:dsetup} from our data with the following protocol: (1) we measure the setup birefringence (Eq.~\ref{eq:setup}) by performing a polarization rotation measurement without thermal modulation well above the sample's magnetic transition temperature (Fig. \ref{fig:background_sub}a); (2)  we measure $ \partial_t r $ as a function of temperature (Fig. \ref{fig:background_sub}b); (3) we calculate the setup contribution at each temperature by multiplying these quantities, according to Eq.~\ref{eq:dsetup} (Fig. \ref{fig:background_sub}c, \ref{fig:background_sub}d); (4) finally, we subtract the setup contribution from the thermally-modulated signal measured as a function of temperature. This leaves us with just the contribution from the sample, as demonstrated for example in Ref.~\cite{ye_elastocaloric_2023}.

\subsection{Obtaining sample parameters}

After subtracting the background, the thermally-modulated signal amounts to the derivative of \ref{eq:idealsample} with respect to temperature: 

\begin{equation}
\frac{d\Theta(\varphi, T)}{dT} =
\frac{d\theta_B(T)}{dT} \sin[2(\varphi + \varphi_0(T))] +
2\theta_B(T) \cos[2(\varphi + \varphi_0(T))] \frac{d\varphi_0(T)}{dT} +
\frac{d\theta_K(T)}{dT}.
\label{eq:tempderivative}
\end{equation}

\noindent The thermally-modulated polarization rotation experiments have both a sinusoidal and a polarization-independent component. The polarization-independent $ \frac{d\theta_K(T)}{dT} $ term is straightforward to interpret: it is a temperature derivative of MOKE. Since no further analysis is required for the MOKE signal, and since it is completely separable from birefringence, we omit it from the remaining analysis.

The observation of the sinusoidal component in \ref{eq:tempderivative} (first two terms) immediately proves a non-zero birefringence ($\theta_B\neq0$), but interpreting its magnitude and angular dependence requires more analysis, since the amplitude $ \theta_B $ and orientation $ \phi_0 $, as well as their temperature derivatives, affect the measured signal. Both sinusoidally varying terms in \ref{eq:tempderivative} have the same periodicity, so we can rewrite them as: 

\begin{equation}
\frac{d\Theta(\varphi, T)}{dT} = A(T) \cos[2(\varphi + \varphi_A(T))]=A(T) \left[ \cos\big(2\varphi_A(T)\big)\cos\big(2\varphi\big) - \sin\big(2\varphi_A(T)\big)\sin\big(2\varphi\big) \right]
\label{eq:fit}
\end{equation}

\noindent where $A(T)$ and $ \varphi_A(T)$ now represent the amplitude and the orientation of thermally-modulated birefringence, respectively.

Integrating Eq.~\ref{eq:fit}, we find:

\begin{align}
    \Theta(\varphi, T) &= 
    \cos\bigl(2\varphi\bigr) \int_{T \gg T_N}^{T} \! A(T') \cos\bigl(2\varphi_A(T')\bigr) \, dT'  - \sin\bigl(2\varphi\bigr) \int_{T \gg T_N}^{T} \! A(T') \sin\bigl(2\varphi_A(T')\bigr) \, dT' \notag \\
    &= A_c \cos\bigl(2\varphi\bigr) + A_s \sin\bigl(2\varphi\bigr),
\end{align}
where we used the fact that $\Theta(\varphi, T>T_N) = 0$. In the last line we defined $A_c$ and $ A_s$, which can both be calculated from the measured quantities $A(T)$ and $\varphi_A(T)$.

Comparing to our original expression for the sample-induced polarization rotation (Eq.~\ref{eq:idealsample}), we can obtain the $\theta_B$ and $\phi_0$ from the measured quantities as:

\begin{equation}
\begin{aligned}
\theta_B = \sqrt{A_c^2 + A_s^2} \\
\varphi_0 = \frac{1}{2} \arctan\left( \frac{A_c}{A_s} \right)
\end{aligned}
\end{equation}

In Fig. 2b of the main text we show the temperature derivative of $ \theta_B $ for better comparison with the measured MOKE signal, $\frac{d\theta_K(T)}{dT} $. We also note that our full fitting function used for rotation measurements includes a $ 4\varphi $ dependent term:

\begin{equation}
\frac{d\Theta(\varphi, T)}{dT} = A \cos[2(\varphi + \varphi_A(T))] + 
B \cos[4(\varphi + \varphi_{B}(T))] + \frac{d\theta_K(T)}{dT}
\label{eq:fit2}
\end{equation}

\noindent to account for any possible contribution from the setup retardance. This contribution is negligible after the previously described background subtraction, as is observed in Fig. 1f of the main text.

\section{Symmetry analysis}

Below we describe in detail the symmetry analysis whose results are shown in the main text. As described in the main text, \underdoped~ and \ideal~ behave very differently. The ordering vector determined from neutron scattering is different, while the crystal structure, as measured by x-ray scattering, appears mostly unchanged~\cite{park_composition_2024}. We will refer to the two samples as $x=0.32$ and $x=1/3$. We begin our symmetry analysis with the usual assumption that the crystal space group of $x=0.32$ is the same as that of $x=1/3$ (no. 182). Using that starting point, we will utilize a symmetry-based representation analysis to determine magnetic structures consistent with the experimental data on \underdoped~for both magnetic phases. 

\subsection{Identifying the irreps}

In \underdoped~ there are two phase transitions and two distinct ordered magnetic phases:
\begin{itemize}
    \item Higher temperature phase (HT): Long range magnetic order with spins parallel to $c$. Peaks in neutron scattering are visible at $\bm{q}_{HT}=(0,1/2,0)$ and $\bm{q}'_{HT}=(1/2,1/2,0)$~\cite{takagi_spontaneous_2023}. 
    \item Lower temperature phase (LT): The in-plane spins ($\perp c$) develop a long range order. The intensity of the $\bm{q}_{HT}=(0,1/2,0)$ peak varies smoothly through the low temperature transition, while the intensity of the $\bm{q}'_{HT}=(1/2,1/2,0)$ peak acquires an additional contribution~\cite{takagi_spontaneous_2023}. We conclude that the  spin order $\parallel c$  remains unaffected by the lower temperature transition, while an additional order parameter, with spins $\perp c$ develops. 
\end{itemize}

Next, we use Isotropy~\cite{stokes_isotropy} to find all possible order parameters consistent with the scattering patterns and the parent space group of \ideal (no. 182). Noting that all the observed $q$-values correspond to (the star of) the $M$ point in $k$-space, we run the following code in Isotropy: 

\begin{verbatim}
VALUE PARENT 182
VALUE KPOINT M
VALUE WYCKOFF C
SHOW IRREP 
SHOW MICROSCOPIC VECTOR PSEUDO
DISPLAY DISTORTION
\end{verbatim}

\noindent This returns a list of all the magnetic irreducible representations (irreps) compatible with the parent group, which we reproduce here in Tables~\ref{tab:out-of-plane-structures} and \ref{tab:in-plane-structures} for out-of-plane and in-plane spins, respectively. We note that all the irreps are three-dimensional: all the candidate order parameters have three components that are related by 3-fold rotational symmetry. The order parameters are vectors in three-dimensional spaces spanned by those components, so multiple magnetic structures belong to each irrep. For example, some order parameters contain only one component (1\textbf{Q} structure), some correspond to coherent superpositions of all three (3\textbf{Q} structure), while others are in between. First, we determine which irreps are consistent with the scattering patterns.

\begingroup
\scriptsize               % smaller font (try \footnotesize if you want slightly larger)
\renewcommand{\arraystretch}{0.75} % tighter vertical spacing
\begin{table}[h!]
\centering
\begin{tabular}{@{}llccc@{}}
\toprule
\textbf{Irrep} & \textbf{Co coordinate} & \multicolumn{3}{c}{\textbf{Magnetic moment}} \\
\cmidrule(lr){3-5}
 & & \textbf{Component 1} & \textbf{Component 2} & \textbf{Component 3} \\
\midrule
\multirow{8}{*}{$\mathrm{mM}_2$} & (1/3,2/3,1/4) & (0,0,1) & (0,0,-1) & (0,0,1) \\
 & (1/3,5/3,1/4) & (0,0,1) & (0,0,1) & (0,0,-1) \\
 & (4/3,2/3,1/4) & (0,0,-1) & (0,0,-1) & (0,0,-1) \\
 & (4/3,5/3,1/4) & (0,0,-1) & (0,0,1) & (0,0,1) \\
 & (-1/3,1/3,3/4) & (0,0,1) & (0,0,1) & (0,0,-1) \\
 & (-1/3,4/3,3/4) & (0,0,1) & (0,0,-1) & (0,0,1) \\
 & (2/3,1/3,3/4) & (0,0,-1) & (0,0,1) & (0,0,1) \\
 & (2/3,4/3,3/4) & (0,0,-1) & (0,0,-1) & (0,0,-1) \\
\midrule
\multirow{8}{*}{$\mathrm{mM}_3$} & (1/3,2/3,1/4) & (0,0,1) & (0,0,1) & (0,0,1) \\
 & (1/3,5/3,1/4) & (0,0,1) & (0,0,-1) & (0,0,-1) \\
 & (4/3,2/3,1/4) & (0,0,-1) & (0,0,1) & (0,0,-1) \\
 & (4/3,5/3,1/4) & (0,0,-1) & (0,0,-1) & (0,0,1) \\
 & (-1/3,1/3,3/4) & (0,0,-1) & (0,0,1) & (0,0,1) \\
 & (-1/3,4/3,3/4) & (0,0,-1) & (0,0,-1) & (0,0,-1) \\
 & (2/3,1/3,3/4) & (0,0,1) & (0,0,1) & (0,0,-1) \\
 & (2/3,4/3,3/4) & (0,0,1) & (0,0,-1) & (0,0,1) \\

\bottomrule
\end{tabular}
\caption{Magnetic structures compatible with the parent space group (no. 182), the observed propagation vectors, and out-of-plane spins. The orientation of magnetic moments is given for each Co ion in the magnetic unit cell. Note that for each irrep there are three `components', related by three-fold rotational symmetry.}
\label{tab:out-of-plane-structures}
\end{table}
\endgroup

\begingroup
\scriptsize               % smaller font (try \footnotesize if you want slightly larger)
\renewcommand{\arraystretch}{0.70} % tighter vertical spacing
\begin{table}[h!]
\centering
\begin{tabular}{@{}llccc@{}}
\toprule
\textbf{Irrep} & \textbf{Co coordinate} & \multicolumn{3}{c}{\textbf{Magnetic moment}} \\
\cmidrule(lr){3-5}
 & & \textbf{Component 1} & \textbf{Component 2} & \textbf{Component 3} \\
\midrule
\multirow{8}{*}{$\mathrm{mM}_1$} & (1/3,2/3,1/4) & (1.155,0.577,0) & (0.577,1.155,0) & (-0.577,0.577,0) \\
 & (1/3,5/3,1/4) & (1.155,0.577,0) & (-0.577,-1.155,0) & (0.577,-0.577,0) \\
 & (4/3,2/3,1/4) & (-1.155,-0.577,0) & (0.577,1.155,0) & (0.577,-0.577,0) \\
 & (4/3,5/3,1/4) & (-1.155,-0.577,0) & (-0.577,-1.155,0) & (-0.577,0.577,0) \\
 & (-1/3,1/3,3/4) & (-1.155,-0.577,0) & (0.577,1.155,0) & (-0.577,0.577,0) \\
 & (-1/3,4/3,3/4) & (-1.155,-0.577,0) & (-0.577,-1.155,0) & (0.577,-0.577,0) \\
 & (2/3,1/3,3/4) & (1.155,0.577,0) & (0.577,1.155,0) & (0.577,-0.577,0) \\
 & (2/3,4/3,3/4) & (1.155,0.577,0) & (-0.577,-1.155,0) & (-0.577,0.577,0) \\
\midrule
\multirow{8}{*}{$\mathrm{mM}_2$} & (1/3,2/3,1/4) & (0,1,0) & (-1,0,0) & (-1,-1,0) \\
 & (1/3,5/3,1/4) & (0,1,0) & (1,0,0) & (1,1,0) \\
 & (4/3,2/3,1/4) & (0,-1,0) & (-1,0,0) & (1,1,0) \\
 & (4/3,5/3,1/4) & (0,-1,0) & (1,0,0) & (-1,-1,0) \\
 & (-1/3,1/3,3/4) & (0,-1,0) & (-1,0,0) & (-1,-1,0) \\
 & (-1/3,4/3,3/4) & (0,-1,0) & (1,0,0) & (1,1,0) \\
 & (2/3,1/3,3/4) & (0,1,0) & (-1,0,0) & (1,1,0) \\
 & (2/3,4/3,3/4) & (0,1,0) & (1,0,0) & (-1,-1,0) \\
\midrule
\multirow{8}{*}{$\mathrm{mM}_3$} & (1/3,2/3,1/4) & (0,1,0) & (1,0,0) & (-1,-1,0) \\
 & (1/3,5/3,1/4) & (0,1,0) & (-1,0,0) & (1,1,0) \\
 & (4/3,2/3,1/4) & (0,-1,0) & (1,0,0) & (1,1,0) \\
 & (4/3,5/3,1/4) & (0,-1,0) & (-1,0,0) & (-1,-1,0) \\
 & (-1/3,1/3,3/4) & (0,1,0) & (-1,0,0) & (1,1,0) \\
 & (-1/3,4/3,3/4) & (0,1,0) & (1,0,0) & (-1,-1,0) \\
 & (2/3,1/3,3/4) & (0,-1,0) & (-1,0,0) & (-1,-1,0) \\
 & (2/3,4/3,3/4) & (0,-1,0) & (1,0,0) & (1,1,0) \\
\midrule
\multirow{8}{*}{$\mathrm{mM}_4$} & (1/3,2/3,1/4) & (1.155,0.577,0) & (-0.577,-1.155,0) & (-0.577,0.577,0) \\
 & (1/3,5/3,1/4) & (1.155,0.577,0) & (0.577,1.155,0) & (0.577,-0.577,0) \\
 & (4/3,2/3,1/4) & (-1.155,-0.577,0) & (-0.577,-1.155,0) & (0.577,-0.577,0) \\
 & (4/3,5/3,1/4) & (-1.155,-0.577,0) & (0.577,1.155,0) & (-0.577,0.577,0) \\
 & (-1/3,1/3,3/4) & (1.155,0.577,0) & (0.577,1.155,0) & (0.577,-0.577,0) \\
 & (-1/3,4/3,3/4) & (1.155,0.577,0) & (-0.577,-1.155,0) & (-0.577,0.577,0) \\
 & (2/3,1/3,3/4) & (-1.155,-0.577,0) & (0.577,1.155,0) & (-0.577,0.577,0) \\
 & (2/3,4/3,3/4) & (-1.155,-0.577,0) & (-0.577,-1.155,0) & (0.577,-0.577,0) \\
\bottomrule
\end{tabular}
\caption{Magnetic irreps compatible with the parent space group (no. 182), the observed propagation vectors, and in-plane spins. The orientation of magnetic moments is given for each Co ion in the magnetic unit cell. Note that for each structure there are three `components', related by three-fold rotation symmetry.}
\label{tab:in-plane-structures}
\end{table}

\endgroup

\newpage
\subsubsection{Out-of-plane spins}

\begin{figure}
    \centering
    \includegraphics[width=1\linewidth]{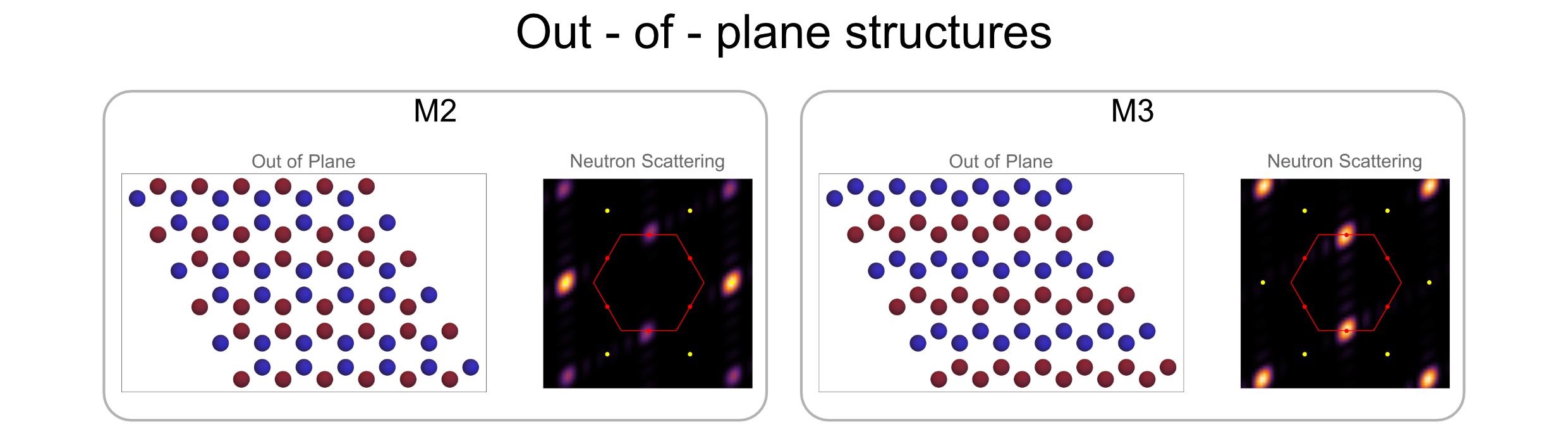}
    \caption{Out-of-plane spins. For each irrep $\mathrm{mM}_2$ and $\mathrm{mM}_3$, we show (a) a top view of the lattice of Co ions, and magnetic order corresponding to component 2 of the order parameter belonging to the irrep (\textit{cf.} Table~\ref{tab:out-of-plane-structures}) with the red and blue colors representing the spin up and spin down; (b) the neutron scattering intensity corresponding to the structure in (a). }
    \label{fig:OutOfPlaneSpins}
\end{figure}

There are two magnetic irreps consistent with out-of-plane spins, $\bm{q}_{0.32}=0.5$ and the parent space group:  $\mathrm{mM}_2$ and $\mathrm{mM}_3$ (Table~\ref{tab:out-of-plane-structures}). In Fig.~\ref{fig:OutOfPlaneSpins} we plot the spin arrangement and the neutron scattering intensity for component 2 of each of them (components 1 and 3 are obtained by applying a 3-fold rotation). We use the spin orientations in Table~\ref{tab:out-of-plane-structures} and lattice vectors to obtain the real space picture. To get the neutron scattering intensity we multiply the Fourier transform of the real space spin density with $\bm{S} \perp \bm{q}$, where $\bm{q}$ is the momentum transfer.

For $\mathrm{mM}_2$ the scattering intensity is observed both at $\bm{q}=(0,1/2,0)$ and at $\bm{q}=(1/2,1/2,0)$, while for M3 it is seen only at $\bm{q}=(0,1/2,0)$. The former is consistent with the experiment, confirming $\mathrm{mM}_2$ as the irrep corresponding to the out-of-plane order.

\newpage
\subsubsection{In-plane spins}

\begin{figure}[h]
    \centering
    \includegraphics[width=1\linewidth]{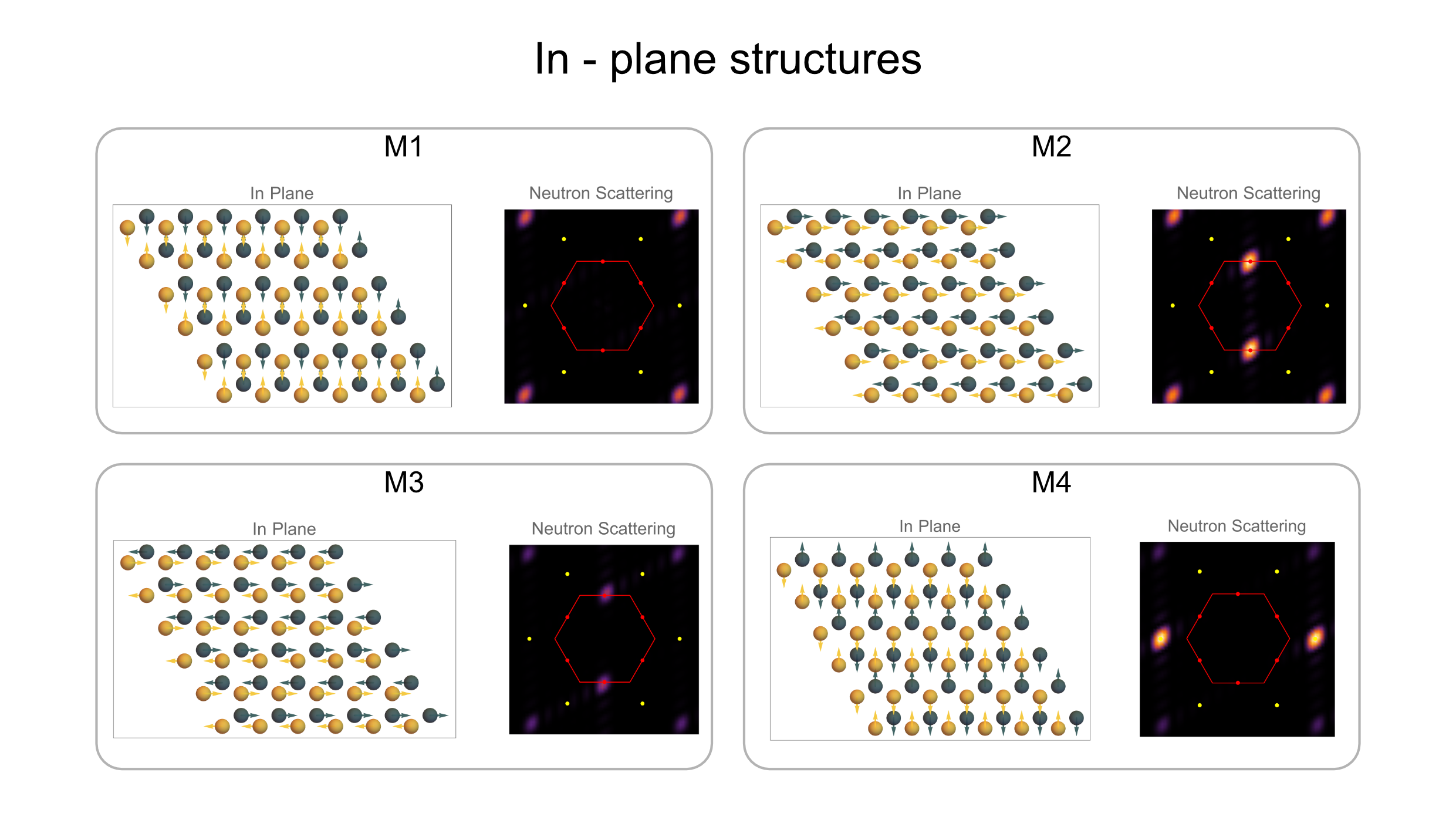}
    \caption{In-plane spins. For each irrep $\mathrm{mM}_1$ - $\mathrm{mM}_4$, we show (a) the lattice of Co ions, with arrows representing the in-plane spin orientations corresponding to the component 2 of the order parameter belonging to the irrep (\textit{cf.} Table~\ref{tab:in-plane-structures}), and the color (yellow/green) encoding the $z$ - position of the atom; (b) the neutron scattering intensity. The neutron scattering pattern and the Fourier transform are not necessarily the same, since neutron scattering cross sections suppress intensity for $\bm{S} \parallel \bm{q}$.}
    \label{fig:InPlaneSpins}
\end{figure}

We repeat the procedure for in-plane spins (Fig~\ref{fig:InPlaneSpins}, Table~\ref{tab:in-plane-structures}). There are now four possible irreps,  and we find that only $\mathrm{mM}_4$ is consistent with the observed scattering pattern.

\newpage

\subsection{Identifying structures compatible with optics}

\subsubsection{High temperature phase}\label{sec:HTgrouplist}
%\subsubsection{List of magnetic space groups}\label{sec:HTgrouplist}
Now that we know the relevant irreps, we can identify the magnetic structures, starting with the high temperature state. Again, we use Isotropy~\cite{stokes_isotropy}: 
\begin{verbatim}
SET MAGNETIC
VALUE PARENT 182
SETTING MILLER-LOVE
VALUE IRREP M2
SHOW SUBGROUP
SHOW BASIS
SHOW ORIGIN
SHOW DIRECTION VECTOR
DISPLAY ISOTROPY
\end{verbatim}

\noindent Note that we had to explicitly state that the calculation is magnetic. The output,  shown in Table~\ref{tab:subgroups}, lists all the magnetic space groups that can be associated with the irrep $\mathrm{mM}_2$. The basis vectors and the origin refer to the definition of the coordinate system for the space group. The symbols a,b,c, in the `amplitudes' column represent arbitrary real numbers, and correspond to the amplitudes of the three order parameter components (Table~\ref{tab:out-of-plane-structures}). For example, the group 19.29 is obtained if the order parameter contains only one component (1\textbf{Q} structure), while 182.183 is obtained through an equal superposition of the three (3\textbf{Q} structure). For each space group there are multiple domains, which we do not list here. 

\begin{table}[h!]
\centering
\begin{tabular}{@{}ccccc@{}}
\toprule
\textbf{Subgroup} & \textbf{OP amplitudes} & \textbf{Basis Vectors}         & \textbf{Origin} \\ 
\midrule
19.29 P2\(_1\)2\(_1\)2\(_1.1'\)\(_C\)[C222\(_1\)] &  (a; 0; 0) & (2, 1, 0), (0, 1, 0), (0, 0, 1)  & (1/2, 1/4, 1/4) \\
20.36 C222\(_1.1'\)\(_a\)[P222\(_1\)]             &  (a; a; 0) & (2, 2, 0), (-2, 2, 0), (0, 0, 1) & (1, 0, 0)       \\
182.183 P6\(_3\)2\('\)2\('\)                      &  (a; a; a) & (0, -2, 0), (2, 2, 0), (0, 0, 1) & (0, 0, 0)       \\
4.10 P2\(_1.1'\)\(_a\)[P2\(_1\)]                  &  (a; b; 0) & (2, 2, 0), (0, 0, 1), (2, 0, 0)  & (0, 0, 0)       \\
20.33 C2\('\)2\('\)2\(_1\)                        &  (a; b; a) & (2, 0, 0), (2, 4, 0), (0, 0, 1)  & (0, 0, 0)       \\
4.7 P2\(_1.1\)                                    &  (a; b; c) & (-2, 0, 0), (0, 0, 1), (0, 2, 0) & (0, 0, 0)       \\
\bottomrule
\end{tabular}
\caption{Subgroups consistent with the $\mathrm{mM}_2$ irrep and out-of-plane spins, directions of order parameters (OP), the amplitudes of the three OP components, basis vectors, and origins for the space group.}
\label{tab:subgroups}
\end{table}

The next step is to constrain the list of possible space groups further using optics. In particular, we use the fact that birefringence is observed in the HT phase, but the anomalous Hall effect (AHE) and MOKE are not (the latter two exhibit the same symmetry constraints within the dipolar approximation for the optical response). 

\begin{table}[h!]
\centering
\begin{tabular}{@{}ccccc@{}}
\toprule
\textbf{Subgroup} & \textbf{OP amplitudes} & \textbf{Birefringence (\checkmark)} & \textbf{MOKE, AHE (\(\times\)  )} \\ 
\midrule
\textbf{19.29}   & (a; 0; 0) & \checkmark &   \(\times\)       \\ 
\textbf{20.36}   & (a; a; 0) & \checkmark &    \(\times\)     \\ 
182.183 & (a; a; a) & \(\times\) &  \checkmark        \\ 
\textbf{4.10}    & (a; b; 0) & \checkmark &   \(\times\)       \\ 
20.33   & (a; b; a) & \checkmark &  \checkmark        \\ 
4.7     & (a; b; c) & \checkmark &  \checkmark        \\ 
\bottomrule
\end{tabular}
\caption{Table showing subgroups with OP amplitudes, and whether birefringence  and MOKE are allowed by symmetries of the group: \checkmark indicates they are allowed, and \(\times\) that they are not. The experimental finding is indicated in the table heading. Subgroups whose labels are shown in bold are compatible with experimental findings.}
\label{tab:birefringence_kerr_op_amplitudes}
\end{table}

Birefringence is allowed by all the space groups which do not have a rotational symmetry $R_n$ with $n>2$. Therefore, only the equal superposition of three order parameter components (group 182.183) prohibits it. To find which groups are compatible with MOKE, we use the MTENSOR tool from the Bilbao crystallographic server~\cite{Gallego:lk5043}. The input for the tool is the magnetic space group, and the outputs are the symmetry-allowed forms of tensors of interest; the spontaneous Faraday effect in this case (same symmetry as MOKE and AHE). 

We identify three structures for the HT phase which are compatible with the observation of birefringence and no MOKE: a 1\textbf{Q} structure (19.29), and two 2\textbf{Q} structures, with equal (20.36) and different (4.10) amplitudes of the two components. The 3\textbf{Q} structure (182.183) is not compatible with either finding, as it would prohibit birefringence and allow MOKE (Table~\ref{tab:birefringence_kerr_op_amplitudes}).

\subsubsection{Low temperature phase}

%\subsubsection{List of magnetic space groups}\label{sec:LTgrouplist}
We repeat the procedure of Sec.~\ref{sec:HTgrouplist} for the low temperature phase. An important difference is that we now have two coupled order parameters which correspond to two irreps: $\mathrm{mM}_2$ and $\mathrm{mM}_4$. The relevant Isotropy~\cite{stokes_isotropy} commands are given below, and the list of space groups is in Table~\ref{tab:subgroups_detailed}.

\begin{verbatim}
SET MAGNETIC
VALUE PARENT 182
SETTING MILLER-LOVE
VALUE IRREP M2 M4
SHOW SUBGROUP
SHOW BASIS
SHOW ORIGIN
SHOW DIRECTION VECTOR
DISPLAY ISOTROPY COUPLED
\end{verbatim}

\begin{table}[h!]
\centering
\begin{tabular}{@{}ccccc@{}}
\toprule
\textbf{Subgroup} & \textbf{OP amplitudes ($\mathrm{mM}_2$ \(\mid\) $\mathrm{mM}_4$)}           & \textbf{Basis Vectors}               & \textbf{Origin} \\ 
\midrule
4.12 P\textsubscript{c}2\textsubscript{1} & (a; 0; 0 \(\mid\) b; 0; 0) & (2, 1, 0), (0, 1, 0), (0, 0, 1) & (1/2, 1/4, 0) \\
4.10 P2\(_1.1'\)\(_a\)[P2\(_1\)]  & (a; 0; 0 \(\mid\) 0; b; 0)   & (2, 2, 0), (0, 0, 1), (2, 0, 0)      & (1, 1/2, 0)     \\
20.34 C22\('\)2\(_1'\)          & (a; a; 0 \(\mid\) 0; 0; b)   & (-2, 2, 0), (-2, -2, 0), (0, 0, 1)   & (-1/2, 3/2, 1/4) \\
4.9 P2\(_1'\)                & (a; b; 0 \(\mid\) 0; 0; c)   & (-2, 0, 0), (0, 0, 1), (0, 2, 0)     & (-1/2, 1/2, 0)  \\
20.33 C2\('\)2\('\)2\(_1\)    & (a; 0; 0 \(\mid\) 0; b; b)   & (0, 2, 0), (-4, -2, 0), (0, 0, 1)    & (1, 3/2, 0)     \\
5.17 C2.1\(_a\)[P2]           & (a; a; 0 \(\mid\) b; b; 0)   & (2, 2, 0), (-2, 2, 0), (0, 0, 1)     & (0, 1, 1/4)     \\
5.17 C2.1\(_a\)[P2]         & (a; a; 0 \(\mid\) b; -b; 0)  & (2, -2, 0), (2, 2, 0), (0, 0, 1)     & (1, 0, 0)       \\
5.15 C2\('\)               & (a; b; a \(\mid\) -c; 0; c)  & (2, 0, 0), (2, 4, 0), (0, 0, 1)      & (0, 0, 1/4)     \\
150.27 P32\('\)1         & (a; a; a \(\mid\) -b; b; -b) & (0, -2, 0), (2, 2, 0), (0, 0, 1)     & (0, 0, 0)       \\
4.7 P2\(_1.1\)          & (a; 0; 0 \(\mid\) 0; b; c)   & (-2, 0, 0), (0, 0, 1), (0, 2, 0)     & (0, 1/2, 0)     \\
1.3 P1.1\('\)\textsubscript{c}[P1]           & (a; b; 0 \(\mid\) c; d; 0)   & (0, 0, 1), (2, 0, 0), (2, 2, 0)      & (0, 0, 0)       \\
5.13 C2.1                 & (a; a; 0 \(\mid\) b; b; -c)  & (2, 2, 0), (-2, 2, 0), (0, 0, 1)     & (0, 1, 1/4)     \\
5.15 C2\('\)                  & (a; b; a \(\mid\) c; d; c)   & (-2, -4, 0), (2, 0, 0), (0, 0, 1)    & (0, 0, 0)       \\
1.1 P1.1                  & (a; b; c \(\mid\) d; e; f)   & (0, 0, 1), (0, 2, 0), (-2, 0, 0)     & (0, 0, 0)       \\
\bottomrule
\end{tabular}
\caption{Subgroups consistent with the $\mathrm{mM}_2$ (out-of-plane spins) and $\mathrm{mM}_4$ (in-plane spins) coupled irreps, directions of order parameters (OP), the amplitudes of the three OP components, basis vectors, and origins for the space group.}
\label{tab:subgroups_detailed}
\end{table}

%\subsubsection*{Constraints imposed by optics}
Just as we did for the HT phase, we now constrain the order parameters in the LT phase using our optics results (Table~\ref{tab:birefringence_kerr_lmb_op_amplitudes}). We have an additional piece of information now: the observation of the anomalous magneto-birefringence (AMB). AMB is the intrinsic version of the Linear Magneto-Birefringence (LMB)~\cite{eremenko_birefringence_1980,kharchenko_linear_1994,sunko_spin-carrier_2023,donoway_multimodal_2024}, which has two forms, diagonal and off-diagonal. Diagonal LMB refers to the $H$-linear change in birefringence magnitude, and off-diagonal LMB refers to the $H$-linear change in birefringence orientation. Since we observe the change of birefringence orientation whose sign depends on the sign of applied $H_z$, our effect has the same symmetry as off-diagonal LMB. 

To check which magnetic point groups allow AMB, we utilize the fact that LMB, and therefore AMB, has the same symmetry as the piezomagnetic effect: $\varepsilon_{ij}=\Lambda_{ijk}H_k$, where $\varepsilon_{ij}$ is the strain tensor, $H_k$ the applied magnetic field and $\Lambda_{ijk}$ the piezomagnetic tensor. Off-diagonal AMB is therefore allowed in materials that allow $\Lambda_{xyz}$, which can be verified using MAGNDATA ~\cite{Gallego:lk5043}. 

Remarkably, this procedure limits the fourteen magnetic structures compatible with neutron scattering to only three, 4.7, 5.13 and 1.1 (shown in bold in Table~\ref{tab:birefringence_kerr_lmb_op_amplitudes}). If we had not observed AMB, but only MOKE and birefringence, we would be left with a total of seven structures (the additional structures are shown in italic in Table~\ref{tab:birefringence_kerr_lmb_op_amplitudes}), emphasizing the importance of combining information from distinct symmetry-sensitive measurements.

\begin{table}[h!]
\centering
\begin{tabular}{@{}cccccc@{}}
\toprule
\textbf{Subgroup} & \textbf{OP amplitudes} & \textbf{Bir. (\checkmark)}  & \textbf{AHE (\checkmark)}  & \textbf{Off-diagonal AMB} & \textbf{SSC/(1 - $t_{\perp}$/$t_{\parallel}$)} \\ 
\midrule
4.12   & (a; 0; 0 \(\mid\) b; 0; 0)   & \checkmark & \(\times\)                              &    \(\times\)   &    0                        \\
4.10   & (a; 0; 0 \(\mid\) 0; b; 0)   & \checkmark &  \(\times\)                            &       \(\times\)    &    0                    \\
\textit{20.34}& (a; a; 0 \(\mid\) 0; 0; b)   & \checkmark & \checkmark                              &  \(\times\)     &    0                  \\
4.9    & (a; b; 0 \(\mid\) 0; 0; c)   & \checkmark &  \(\times\)                           &   \checkmark    &    0                    \\
\textit{20.33}  & (a; 0; 0 \(\mid\) 0; b; b)   & \checkmark &  \checkmark                          &   \(\times\)            &    -ab$^2$                \\
5.17   & (a; a; 0 \(\mid\) b; b; 0)   & \checkmark &  \(\times\)                             &    \(\times\)  &    0                    \\
5.17   & (a; a; 0 \(\mid\) b; -b; 0)   & \checkmark &  \(\times\)                             &    \(\times\)   &    0                   \\
\textit{5.15}   & (a; b; a \(\mid\) -c; 0; c)  & \checkmark &  \checkmark                             &       \(\times\)   & -b c$^2$              \\
150.27 & (a; a; a \(\mid\) -b; b; -b) & \(\times\) &  \checkmark                             &             \(\times\) & 3 a b$^2$          \\
\textbf{4.7}   & (a; 0; 0 \(\mid\) 0; b; c)   & \checkmark &  \checkmark                             &       \checkmark   & -abc              \\
1.3    & (a; b; 0 \(\mid\) c; d; 0)   & \checkmark &  \(\times\)                             &         \(\times\)   & 0              \\
\textbf{5.13}   & (a; a; 0 \(\mid\) b; b; -c)  & \checkmark &  \checkmark                             &             \checkmark  & 0           \\
\textit{5.15}    & (a; b; a \(\mid\) c; d; c)  & \checkmark &  \checkmark                             &      \(\times\)   & bc$^2$ - 2 acd                    \\
\textbf{1.1}    & (a; b; c \(\mid\) d; e; f)   & \checkmark & \checkmark      &     \checkmark  &  -cde + bdf - aef                          \\
\bottomrule
\end{tabular}
\caption{Table showing coupled subgroups with OP amplitudes, and whether birefringence, MOKE and off-diagonal anomalous magneto-birefringence (AMB) are allowed by symmetries of the group: \checkmark indicates they are allowed, and \(\times\) that they are not. The experimental finding is indicated in the table heading. Subgroups whose labels are shown in bold are compatible with experimental findings. The subgroups shown in italic would be allowed as well if there was no information about AMB. The last column shows scalar spin chirality (SSC) for each of the structures, evaluated according to Eq.~\ref{eq:SSC}.}
\label{tab:birefringence_kerr_lmb_op_amplitudes}
\end{table}

\subsection{Scalar Spin Chirality}

In Ref.~\cite{takagi_spontaneous_2023} scalar spin chirality (SSC) associated with the 3\textbf{Q} structure was identified as a possible mechanism behind the AHE. Although the 3\textbf{Q} structure is incompatible with our findings, here we examine whether the structures that are compatible also exhibit SSC. We use the same approach to calculating SSC that was used in Ref.~\cite{takagi_spontaneous_2023}. Briefly, conduction electrons moving through a triangular plaquette $\alpha$ with three non-coplanar spins $\mathbf{S}_i$, $\mathbf{S}_j$, and $\mathbf{S}_k$ acquire a Berry phase, which acts as a fictitious magnetic flux:
\begin{equation}
    \bm{b}_\alpha \propto t_\alpha \, \chi_\alpha \, \mathbf{n}_\alpha,
\end{equation}
where the scalar spin chirality
\begin{equation}
    \chi_\alpha = \mathbf{S}_i \cdot (\mathbf{S}_j \times \mathbf{S}_k)
\end{equation}
encodes the solid angle spanned by the three spins and is nonzero only for non-coplanar configurations. Here, $t_\alpha$ is the electron hopping amplitude along the loop $i \to j \to k \to i$, and $\mathbf{n}_\alpha$ is the unit normal vector to the plaquette. In Co$_x$TaS$_2$ there are two types of plaquettes, with different values of $t_\alpha$: ones comprising of Co atoms in Co planes ($t_\alpha=t_{\parallel}$), and the ones connecting distinct Co planes ($t_\alpha=t_{\perp}$). 

The total SSC-induced fictitious magnetic field, $b$ is given by a sum of $\bm{b}_\alpha$ for all plaquettes in a magnetic unit cell. When evaluated for magnetic structures described by $(a_1,a_2,a_2|b_1,b_2,b_2)$ (Table~\ref{tab:birefringence_kerr_lmb_op_amplitudes}), we find  
\begin{equation}
    \bm{b} \propto \left(1-\frac{t_\perp}{t_\parallel}\right)\left( a_2 b_1 b_3 - a_3 b_1 b_2 - a_1 b_2 b_3 \right) \bm{\hat{z}}.
\end{equation}\label{eq:SSC}
This result is consistent with findings of Ref.~\cite{takagi_spontaneous_2023}: SSC would vanish for $t_\perp=t_\parallel$, and it is non-zero in the 3\textbf{Q} structure ($1=a_1=a_2=a_3=-b_1=b_2=-b_3$, MSG 150.27, Table~\ref{tab:subgroups_detailed}). 

However, the expression for SSC in terms of the $\mathrm{mM}_4$ and $\mathrm{mM}_2$ orders offers an important new insight: $1\bm{Q} +2\bm{Q}$ order, described by $a_ib_jb_k$, with $i$, $j$ and $k$ all different, hosts SSC for all parameter values! Our findings therefore show that SSC is not limited to 3\textbf{Q} structures.

\newpage

\end{document}